\documentclass[sigconf]{acmart}
\usepackage{graphicx}
\usepackage{array}
\usepackage{hyperref}
\usepackage{float}
\usepackage{caption}
\usepackage{multirow}
\usepackage{subcaption}
\usepackage[ruled,vlined]{algorithm2e}

\copyrightyear{2020}
\acmYear{2020}
\setcopyright{acmcopyright}
\acmConference[IndiaHCI 2020]{IndiaHCI '20: Proceedings of the 11th Indian Conference on Human-Computer Interaction}{November 5--8, 2020}{Online, India}
\acmBooktitle{IndiaHCI '20: Proceedings of the 11th Indian Conference on Human-Computer Interaction (IndiaHCI 2020), November 5--8, 2020, Online, India}
\acmPrice{15.00}
\acmDOI{10.1145/3429290.3429307}
\acmISBN{978-1-4503-8944-0/20/11}

\begin{document}

\title{\textit{Hear Her Fear}: Data Sonification for Sensitizing Society 
on Crime Against Women in India 
}


\author{Surabhi S. Nath}
\affiliation{Computer Science and Engineering\\
\institution{IIIT Delhi\\New Delhi, India}}
\email{surabhi16271@iiitd.ac.in}

\renewcommand{\shortauthors}{S. S. Nath}

\begin{abstract}
Data sonification is a means of representing data through sound and has been utilized in a variety of applications. Crime against women has been a rising concern in India. We explore the potential of data sonification to provide an immersive engagement with sensitive data on crime against women in Indian states. The data for nine crime categories covering thirty-five Indian states over a period of twelve years is acquired from National records. Sonification techniques of parameter mapping and auditory icons are adopted: sound parameters such as frequencies, amplitudes and timbres are incorporated to represent the crime data, and audio sounds of women screams are employed as auditory icons to emphasize the traumatic experience. Higher crime rates are assigned higher frequencies, harsher scream textures and larger amplitudes. A user-friendly interface is developed with multiple options for sequential and comparative data sonification. Through the interface, a user can evaluate and compare the extent of crime against women in different states, years or crime categories. Sound spatialization is used to immerse the listener in the sound and further intensify the sonification experience. To assess and validate effectiveness, a user study on twenty participants is conducted with feedback obtained through questionnaires. The responses indicate that the participants could comprehend trends in the data easily and found the data sonification experience impactful. Sonification may therefore prove to be a valuable tool for data representation in fields related to social and human studies.
\end{abstract}




\maketitle

\section{Introduction}
Research today demands appropriate and efficient techniques of data presentation for effective communication. Sonification is a versatile means of using sound for representing data. Sound can prove to be an impactful medium to represent data through its numerous parameters such as frequency, amplitude, timbre that can be adaptively controlled \cite{worrall2019intelligible}. Sound, with its multi-dimensional nature, can be extremely powerful as it has the potential to capture features that may be missed out in a visual representation. Moreover, dynamic data are better understood through sound as its temporal nature can enable meaningful expression. In combination with visuals, sonification is an effective form of data portrayal as it has complementary properties which can enhance visual presentation \cite{hermann2011sonification}. Further, this medium improves accessibility by reaching out to the visually impaired population \cite{ali2019sonify}.

Data sonification has been applied in a variety of fields including medicine, finance, climatology and music composition \cite{kramer2010sonification, kather2017polyphonic, hildebrandt2012applying, dell2016sonification}. Besides the wide spectrum of use-cases, a few studies apply sonification to address social concerns. Lenzi et al. in their review article discuss five different projects applying sonification to socially relevant issues and compare them based on intentionality in their design \cite{lenzi2020intentionality}. Emotional regulation has been attempted in a study through sonification of physiological data of automobile drivers to increase self-awareness and ensure road safety \cite{landry2016listen}. Sonification has also been used to communicate data on alcohol health risks \cite{walus2016sonification}. Furfaro et al. used interactive sonification to measure emotional, perceptual and motor behaviour of individuals \cite{furfaro2015sonification}. Work by Oh et al. used biometric data to extract the degree of sleep and heart rate, which was sonified and displayed as 3D animation for diagnosing sleep disorders \cite{oh2018audiovisual}. Furthermore, Buckley presented a new way of sensitizing students on risks regarding loan debts through musical scores \cite{buckley2019tackling}. Our study deals with another of today's growing societal concerns in the Indian context – crime against women.

Crime against women is seen as an issue of public health and violation of human rights worldwide 
\cite{russo201912}. It is viewed as a serious setback to a country's progress and prevails across income and education levels globally. In India, despite some efforts to secure women’s human rights, the situation continues to be abysmal and challenging \cite{verma2017exploring}. Data on crime against women in India is published by the National Crime Records Bureau (NCRB)\footnote{\href{https://ncrb.gov.in/en/crime-india}{ncrb.gov.in/en/crime-india}}, Government of India. As per the statistics, the total number of crimes against women all India has increased by 70\% from 2001 to 2012. In some states like Assam, Bihar, Jharkhand, Delhi, Odisha, and many North-Eastern States, the number of crimes has more than doubled, with an almost five times increase in West Bengal in the same period. More recently, the MeToo Movement in India has gained momentum with women openly voicing the harassment they face \cite{pegu2019metoo}. To call attention to this situation, the International Centre for Research on Women proposes to use media in creative ways to encourage introspection on the social attitudes and problems of crime against women in the society \cite{ICRW}.

The aim of our study is to explore the potential of data sonification for communicating the rising crime rates in Indian states effectively and sensitizing society on this pressing issue. We have developed a user-friendly sonification interface which enables comparison of the scenarios in various states, crime categories and years, and identifies the critical cases demanding urgent attention. We have tested the efficacy of the designed interface by conducting a preliminary user-study that also employs sound spatialization. To the best of our knowledge, this is the first work on crime against women using data sonification.

\section{Data}
\subsection{Source}
Data on the number of crime cases against women for 35 Indian states (including Union Territories) and All India, over 12 years from 2001 to 2012 on 9 crime categories, including total crimes were obtained from records published by NCRB in the Open Government Data Platform India, under public domain\footnote{\href{https://data.gov.in/resources/crime-against-women-during-2001-2012}{data.gov.in/resources/crime-against-women-during-2001-2012}}. The data fields comprised of State [1...36], Crime Category [1...9] and Year [2001...2012], resulting in a total data size of 36 x 9 x 12. Data for years subsequent to 2012 were not available. The nine crime categories comprise of Rape, Kidnapping \& Abduction, Dowry Deaths, Assault on Women with Intent to Outrage her Modesty, Insult to the Modesty of Women, Cruelty by Husband or Relatives, Immoral Traffic, Indecent Representation of Women and Total Crimes Against Women.

\subsection{Processing}
The available data recorded the absolute number of crime cases under the various heads. To enable meaningful interpretation of the number of crimes, we incorporated the population change over the years. Due to lack of the population data of each year, the decadal percent population growth (2001-2011) in all states published by the Census of India was used\footnote{\href{http://censusindia.gov.in/2011-prov-results/data_files/india/Final_PPT_2011_chapter3.pdf}{censusindia.gov.in/2011-prov-results/datafiles/india}}. The population change across the twelve years was assumed to be uniform. The yearly number of crime cases was proportionately altered to accommodate for this change in each year (Algorithm \ref{algo1}). Further, the crime data was normalized by subtracting the value in the base year 2001 from the value in each year, which shifted the distribution to a starting value of 0 in 2001 (Algorithm \ref{algo2}).

\begin{algorithm}[t]
\LinesNumbered
\SetKwData{Left}{left}\SetKwData{This}{this}\SetKwData{Up}{up}
\SetKwFunction{Union}{Union}\SetKwFunction{FindCompress}{FindCompress}
\SetKwInOut{Input}{Input}\SetKwInOut{Output}{Output}
\Input{Set of states $I$ and set of crime categories $J$}
\BlankLine
{For a state $i \in I$, the decadal percent population growth = $x_i$}\\
{Annual percent population growths in 2002, 2003, ..., 2011, 2012 are taken to be $0.1x_i, 0.2x_i..., x_i, 1.1x_i$}\\
{The number of crime cases for state $i$ in each crime category $j \in J$ is hence reduced by $0.1x_i\%, 0.2x_i\%..., x_i\%, 1.1x_i\%$ to get the population incorporated number of crime cases}\\
{This is repeated for all states in $I$}
\caption{Incorporate Population}
\label{algo1}
\end{algorithm}

\begin{algorithm}[t]
\LinesNumbered
\SetKwData{Left}{left}\SetKwData{This}{this}\SetKwData{Up}{up}
\SetKwFunction{Union}{Union}\SetKwFunction{FindCompress}{FindCompress}
\SetKwInOut{Input}{Input}\SetKwInOut{Output}{Output}
\Input{Set of states $I$, set of crime categories $J$ and set of years $K$}
\BlankLine
{For a state $i \in I$, crime category $j \in J$, number of cases in every year $x_k$ ($k \in K$) with first element $x_{2001} = y$ is normalized as $x_k = x_k - y$}\\
{This is repeated for all states in $I$ and crime categories in $J$}
\caption{Data Normalization}
\label{algo2}
\end{algorithm}

\section{Sonification Interface}

\subsection{Design}
We have used the techniques of Parameter Mapping and Auditory Icons for data sonification.

\begin{enumerate}
    \item Parameter mapping: It is the method of conversion of data values to sound parameters. It is a useful technique for conveying data of multi-dimensional nature. In our study, we have primarily experimented with frequencies, amplitudes and timbres since they best describe the sound and are the easiest to perceive.  Synths, or sound generating units are developed to produce sound, and the parameter values are fed from the data in real-time 
\cite{grond2011parameter}.
    \item Auditory Icons: These are self-explanatory real-life sounds, representative of the physical event being sonified. With a semantic content, they enable easy association and add to the emotional perception of the event 
\cite{gaver1986auditory}. We have acquired women screaming sound effects from YouTube to serve as auditory icons since they are the most natural sounds to characterize the pain and misery of the victims subjected to such crimes. Six freely available unique scream sounds are used with increasing harshness in timbre. These sounds were modified using pitch shifts or amplitude factors to map higher instances of crime with higher frequencies, larger amplitudes, harsher timbres, and vice versa.
\end{enumerate}
The computer interface is developed using Supercollider 
\cite{mccartney2002rethinking}, a free and open-source programming environment used for audio synthesis and algorithmic composition (Figure \ref{i}). Supercollider employs a Client-Server architecture and has flexible GUI systems to allow user interaction. The code is made available on GitHub\footnote{\href{https://github.com/surabhisnath/Data-Sonification-Crime-Against-Women-in-Indian-States}{github/surabhisnath/Data-Sonification-Crime-Against-Women-in-Indian-States}}.


\renewcommand{\thefigure}{1}
\begin{figure*}[!pt]
    \begin{subfigure}{\linewidth}
        \centering
        \frame{\includegraphics[scale=.35]{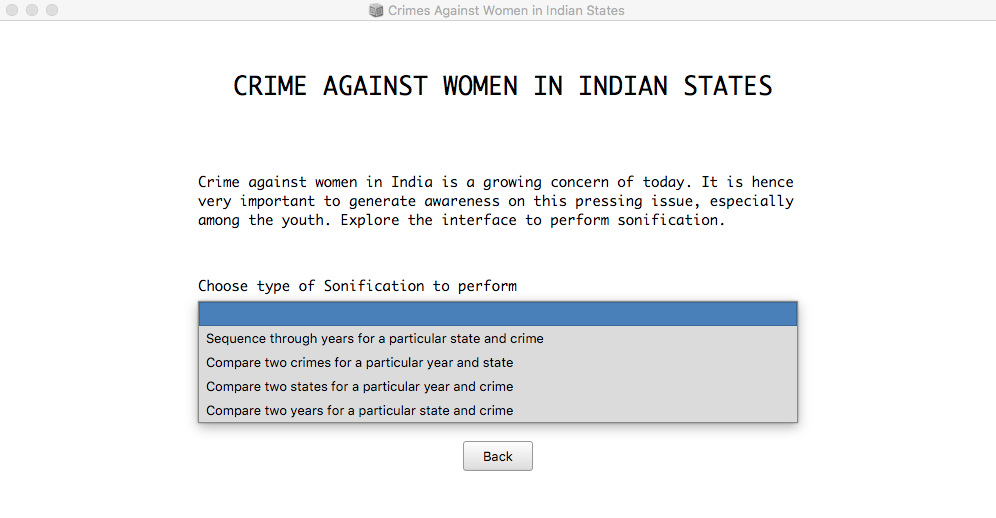}}
        \caption{}
        \label{ione}
    \end{subfigure}
    \\
    \begin{subfigure}{\linewidth}
        \centering
        \frame{\includegraphics[scale=.35]{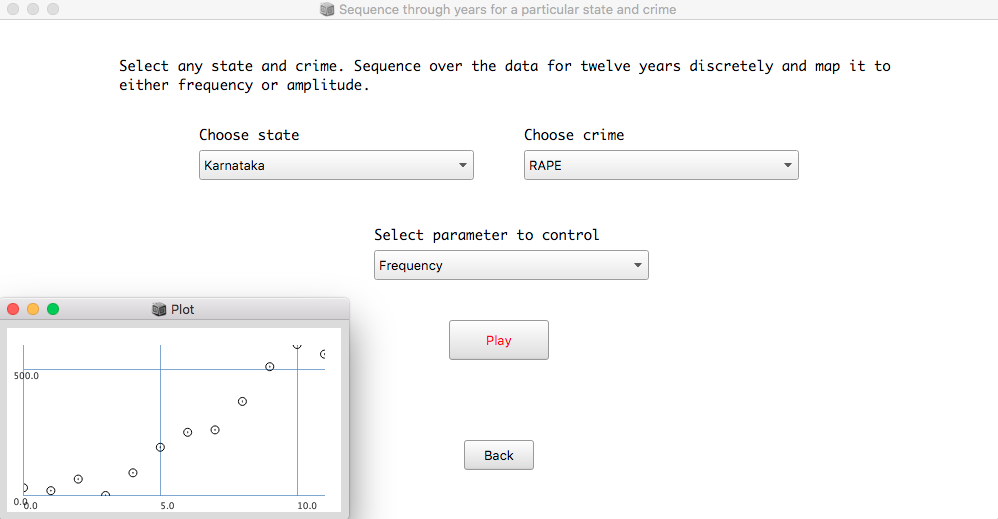}}
        \caption{}
        \label{itwo}
    \end{subfigure}
    \\
    \begin{subfigure}{\linewidth}
        \centering
        \frame{\includegraphics[scale=.351]{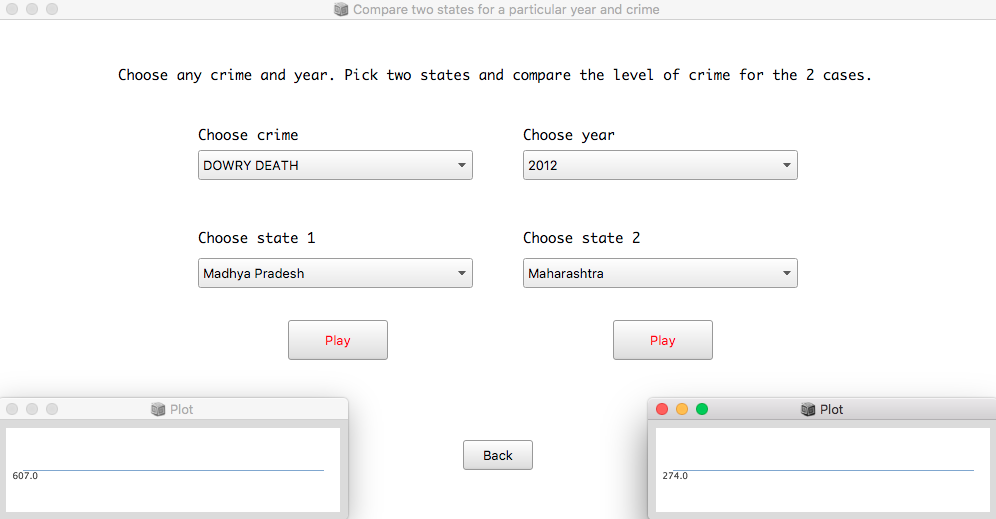}}
        \caption{}
        \label{ithree}
    \end{subfigure}%
\caption{Sonification Interface}
\label{i}
\end{figure*}

\subsection{Output}
The graphical user interface (GUI) opens with a brief introduction to the study and a dropdown to select from four sonification options (Figure \ref{ione}). 


The four options are grouped under two heads – Sequential Data Sonification and Comparative Data Sonification for the purpose of discussion. Every option leads to a new page with instructions for performing the sonification. The GUI provides the sonified output through the "Play" button, and allows switching between pages using the "Back" buttons on each page.

\textbf{Sequential Data Sonification} (Figure \ref{itwo}): Performs sonification of data across the years for a particular state and crime. The user chooses a state and crime category. The crime data values across the twelve years are sonified as screams and played in succession as twelve distinct sounds. The length of each scream is around 1 second. The user has the choice to sonify the crime data as frequencies or as amplitudes. As frequencies, five scream timbres in increasing pitch are selected each mapped to a particular data range. For a higher value of crime data, the scream timbre is harsher and at a higher pitch. As amplitudes, the same baseline frequency scream sound is played in varying loudness based on the crime data values of the twelve years. Higher data values are louder. In both cases, when the twelve sounds are played, the participant can identify the patterns in the data by listening to the variations in the pitch/loudness and timbre of scream sounds. A visual graph is displayed along with the sonified output as feedback for comparison and validation that the data sonification is meaningful.

\textbf{Comparative Data Sonification} (Figure \ref{ithree}): Performs sonification for comparing two crimes for a particular year and state, or for comparing two states for a particular year and crime, or for comparing two years for a particular state and crime. The user can fix any two variables out of state, crime category and year, and select two cases of the third variable for comparison. A single scream sound, sonified based on both frequency and amplitude are played along with the visuals displaying values indicating the number of crimes for the two cases. The scream for the larger data value is louder and at higher pitch. The participant can compare the crime situations in the two cases by differentiating between the two screams.

\section{User Study}

\subsection{Design}
For testing the outcome and evaluating the impact of sonification for the data on crime against women in our study, a user survey was designed for participants to interact with the sonification interface, generate corresponding sonified audio outputs, and interpret the data. User responses were collected through a questionnaire and the effectiveness of sonification was analyzed. The survey was conducted in a Sound Spatialization Lab equipped with 8 speakers for multichannel audio output (Figure \ref{lab}). A two-channel audio output can also be utilized for the study, however this setup was chosen for inducing a more immersive sonic experience.

Twenty participants, a mix of 13 male and 7 female candidates, 19 to 24 years of age, with homogeneous backgrounds, from undergraduate and graduate engineering programs volunteered for the survey. The survey was in accordance with the applicable institute policies. The participants were briefed on the purpose of the study prior to administration. The study was conducted individually for every participant in the lab and was approximately 10 minutes in duration. The door was closed and fans were switched off to prevent other external sound interference. Participants were instructed to test each of the four options available on the interface at least once. They could choose the type of crime, state and year of their interest and could modify the sonification parameters and produce sound output. The participants were undisturbed throughout the session. Soon after, they answered the questionnaire and gave their feedback.

The questionnaire consisted of 12 questions\footnote{\href{https://docs.google.com/forms/d/e/1FAIpQLScz5aICB6xyKKN_JZKBznlLkUTflF36gavtwKnU6hoGrC2Yqg/viewform}{docs.google.com/forms/viewform}}. 4 questions were of multiple-choice type with provision for single option selection. The other 8 questions had a linear rating scale of 1 to 5, with 5 as the most favourable rating. In our analysis, we have considered participant response rating of 4 or 5 to be favourable outcomes.
\renewcommand{\thefigure}{2}
\begin{figure}[H]
    \centering
    \includegraphics[scale=0.06]{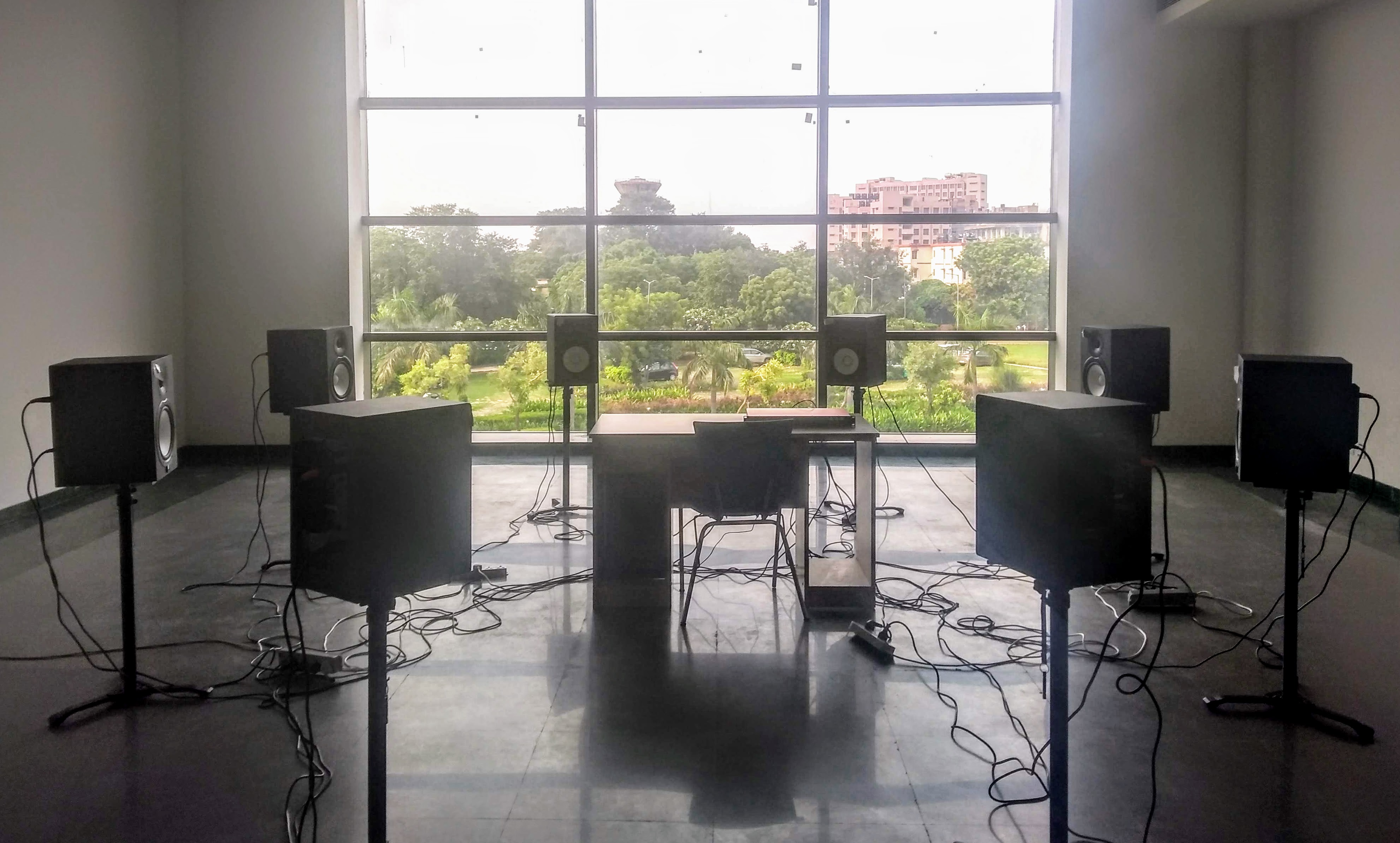}
    \caption{Sound Spatialization Lab}
    \label{lab}
\end{figure}

\subsection{Findings}

\renewcommand{\thefigure}{3}
\begin{figure*}[!pt]
    \begin{subfigure}{0.5\linewidth}
        \centering
        \frame{\includegraphics[scale=.348]{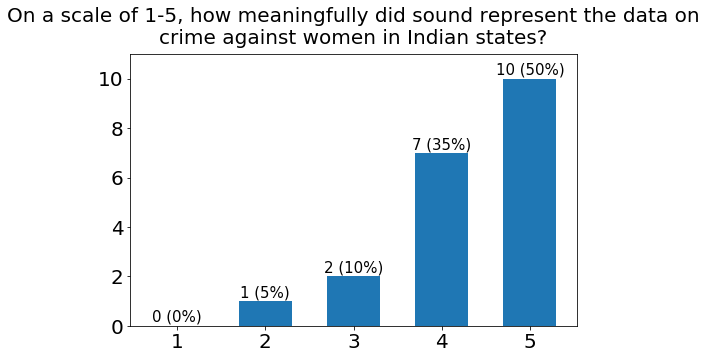}}
        \caption{}
        \label{one}
    \end{subfigure}%
    \begin{subfigure}{0.5\linewidth}
        \centering
        \frame{\includegraphics[scale=.36]{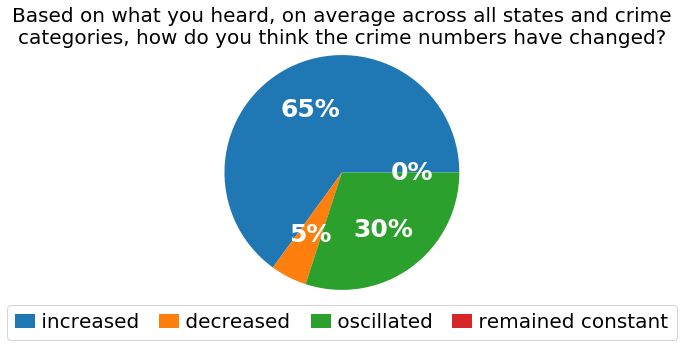}}
        \caption{}
        \label{two}
    \end{subfigure}%
    \\
    \begin{subfigure}{0.5\linewidth}
        \centering
        \frame{\includegraphics[scale=.345]{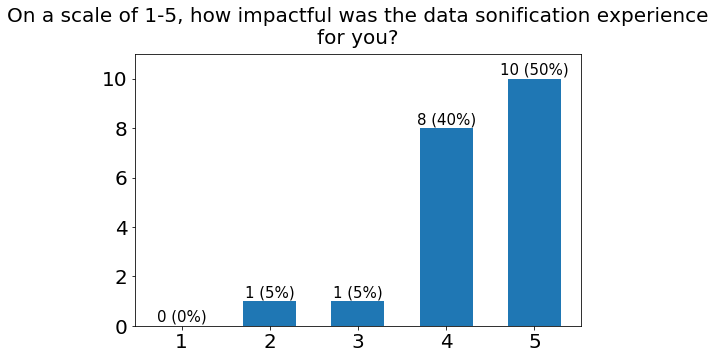}}
        \caption{}
        \label{three}
    \end{subfigure}%
    \begin{subfigure}{0.5\linewidth}
        \centering
        \frame{\includegraphics[scale=.362]{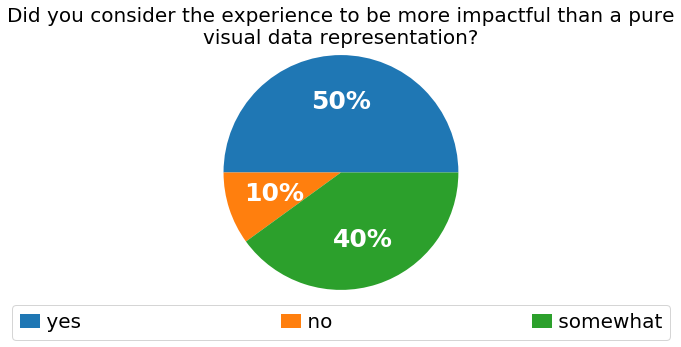}}
        \caption{}
        \label{four}
    \end{subfigure}%
    \\
    \begin{subfigure}{0.5\linewidth}
        \centering
        \frame{\includegraphics[scale=.37]{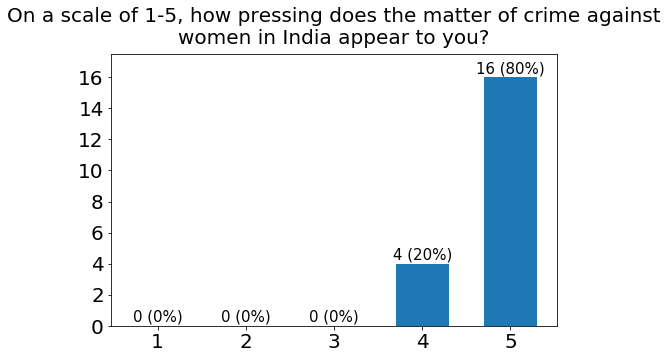}}
        \caption{}
        \label{five}
    \end{subfigure}%
    \begin{subfigure}{0.5\linewidth}
        \centering
        \frame{\includegraphics[scale=.38]{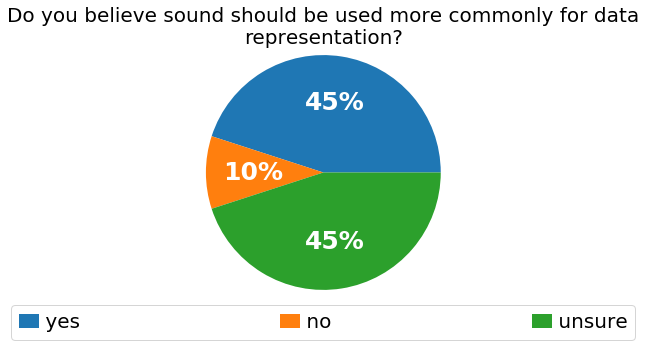}}
        \caption{}
        \label{six}
    \end{subfigure}%
\caption{User Study Findings}
\label{recon}
\end{figure*}

The findings from the participant responses to the questionnaire (Figure \ref{recon}) reveal that data sonification can be an effective medium for representing data of crime against women in Indian states. 85\% of the respondents concurred that the representation was meaningful and that they could understand the varying trends in the data (Figure \ref{one}). While 65\% of the respondents thought that the crime rates in states had increased on average, 30\% said they oscillated across states, crimes or years, and about 5\% thought they decreased with time (Figure \ref{two}). The corresponding underlying data also shows increase, oscillation and reduction in a similar proportion. As sound can easily capture the temporal nature of the data, it may have been easy for respondents to decipher the changing trend in the sonified twelve-year crime data. 

On questions relating to comparing two states, crime categories or years with respect to any one parameter, nearly 95\% of the respondents agreed that they found comparison evident. This demonstrates that the sonification performed was reasonable and difference between the two single screams was easy to comprehend. All the respondents agreed that the crime against women was a pressing issue (Figure \ref{five}) and 90\% reported that the data sonification was highly impactful (Figure \ref{three}). On comparing sound with visual representation, while 50\% respondents thought that the experience was more impactful, 40\% were somewhat unsure and 10\% thought otherwise (Figure \ref{four}). Considering that audio representation is an infrequently used technique and not as mainstream as visuals, the response distribution is still very encouraging. The interface was seen to be easy and self-explanatory by most respondents. Use of sound for data representation in general was supported by about half the number of participants, while the other half were mostly unsure (Figure \ref{six}). This is also understandable given that not all data can be semantically represented through sound.

\section{Limitations}
Although the findings of this study are promising, the work has multiple limitations which would be addressed in future extensions. This preliminary user study is based on a small participant set and hence making definitive conclusions is difficult. The interface design is very basic and has the scope for introducing more features such as pause functionalities, time monitoring, process tracing and additional sound parameters such as tempos, distortions or reverbs to enhance the value of this study. Moreover, open-ended questions or interviews were not included which could add greater understanding of user-interface interaction. Further, ethical concerns such as long-term impact of the sonification experience on the users, particularly on victims or culprits are not addressed. Future experiments could also compare the user experience with and without accompanying visual data display, and evaluate the impact of varied sound spatialization effects.

\section{Conclusion}
The work establishes sound as an effective medium to represent socially relevant data. It demonstrates the potential of data sonification as an immersive user experience to effectively bring out the severity of crime against women in Indian states. It is hoped that these innovations in the presentation of data will make a strong appeal and draw the attention of society to \textit{hear her fear}.

\begin{acks}
The author thanks Prof. Timothy Moyers for his guidance and support, Prof. Grace Eden for her helpful suggestions and IIIT Delhi for providing access to the Sound Spatialization Lab for undertaking this work.
\end{acks}
\bibliographystyle{ACM-Reference-Format}
\bibliography{sample-base}


\end{document}